\documentclass[useAMS,usenatbib]{mn2e}

\usepackage{epsfig}
\def\h2{H$_2$}
\def\f0{$F_0$}

\newcommand\ion[2]{#1$\;${\small\rmfamily\@Roman{#2}}\relax}% 

\onecolumn

\title[Power density spectrum statistics]{Power density spectrum of nonstationary short--lived light curves}
\author[C.~Guidorzi]{Cristiano~Guidorzi\thanks{E-mail:guidorzi@fe.infn.it},
\\
\mbox{}\\
Department of Physics, University of Ferrara, via Saragat 1,  I-44122, Ferrara, Italy
}

\begin{document}

\date{\today}

%\pagerange{\pageref{firstpage}--\pageref{lastpage}} \pubyear{2002}

\maketitle

\label{firstpage}

\begin{abstract}
The power density spectrum of a light curve is often calculated as the average of a number of
spectra derived on individual time intervals the light curve is divided into.
This procedure implicitly assumes that each time interval is a {\em different} sample function
of the {\em same} stochastic ergodic process.
While this assumption can be applied to many astrophysical sources, there remains a class of
transient, highly nonstationary and short--lived events, such as gamma--ray bursts,
for which this approach is often inadequate.
The power spectrum statistics of a constant signal affected by statistical (Poisson) noise
is known to be a $\chi^2_2$ in the Leahy normalisation. However, this is no more the case
when a nonstationary signal is also present.
As a consequence, the uncertainties on the power spectrum cannot be calculated based on the
$\chi^2_2$ properties, as assumed by tools such as XRONOS powspec.
We generalise the result in the case of a nonstationary signal affected by uncorrelated
white noise and show that the new distribution is a non-central $\chi^2_2(\lambda)$,
whose non-central value $\lambda$ is the power spectrum of
the deterministic function describing the nonstationary signal.
Finally, we test these results in the case of synthetic curves of gamma--ray bursts.
We end up with a new formula for calculating the power spectrum uncertainties.
This is crucial in the case of nonstationary short--lived processes affected by
uncorrelated statistical noise, for which ensemble averaging does not make any
physical sense.
\end{abstract}


%%%%%%%%%%%%%%%%%%%%%%%%%%%%%%%%%%%%%%%%%%%%%%
\section{Introduction}
\label{sec:intro}
%%%%%%%%%%%%%%%%%%%%%%%%%%%%%%%%%%%%%%%%%%%%%%
The study of the temporal variability of time series in various branches
of science and engineering has propelled the development of several techniques,
both in frequency and time domains.
Variability studies in the case of astronomical sources are crucial to gain insight
over the dynamical and microphysical timescales, and therefore on the size of the
emitting region as well as the nature itself of the emission process.
This is of key importance in the X-- and $\gamma$--ray domain, where remarkable
flux variations are observed over timescales from days to ms.

Fourier techniques are widely used in this field, as witnessed by the popular
timing analysis package {\sc xronos}\footnote{Available at
{\tt http://heasarc.gsfc.nasa.gov/docs/xanadu/xronos/xronos.html}.} \citep{Stella92}
included in the NASA {\sc heasoft} package\footnote{Available at
{\tt http://heasarc.gsfc.nasa.gov/docs/software/lheasoft/}.}.
The Fourier spectral analysis is fundamental in the study of stationary processes,
since it provides an immediate physical interpretation as a power-frequency
distribution. The Fourier power density spectrum (hereafter, PDS) in particular
decomposes the total variance of a given time series to the different frequencies
thanks to Parseval's theorem (e.g., \citealt{Klis88}; hereafter, K88).
PDS analysis and related tools are suitable to both searching for possible periodic
signals hidden in the data, and to characterising the so-called ``red noise''
connected with the presence of aperiodic variability.

Practically, in the most general case one divides the time series in multiple
adjacent intervals, over which the corresponding PDS is calculated.
Finally, for each frequency bin the resulting PDS value and uncertainty are
the mean and standard deviation of the corresponding power distribution, as is
routinely done by the dedicated {\sc xronos} tool {\tt powspec} \citep{Stella92}.
However, the fundamental assumption behind this procedure is that each time interval
represents a different sampling of the {\em same} stochastic stationary process.
The operation of replacing ensemble averages with time averages of a single
realisation makes sense only if the process is ergodic (e.g., \citealt{Priestley81}).

Astronomical X--ray data rely on photon counting instruments. As such, measurement
uncertainties are often dominated by the photon counting statistics, i.e. the
Poisson distribution. This translates into white noise in the PDS; adopting the
normalisation introduced by \citet{Leahy83} (hereafter L83), the power is known
to follow a $\chi^2$ distribution with 2 degrees of freedom ($\chi^2_2$).

When additional, genuine variability of the source is also present, the resulting
power distribution changes correspondingly. For instance, there is a class of
X--ray sources, such as Cyg~X--1, whose time series are the result of a
process characterised by source--intrinsic correlated noise,
for which time averages of PDSs from individual intervals are still meaningful.
These time series belong to a class of random processes whose PDS is still
compatible with a rescaled $\chi^2_2$ distribution (K88; \citealt{Israel96}).

In the case of highly nonstationary and short--lived events such as
gamma--ray bursts (GRBs), the problem of a proper treatment of the Fourier
PDS requires particular care. Dividing the light curve of a single GRB into
several sub--intervals, and deriving an average PDS does not make any physical
sense, given that the phenomenon is everything but stationary.
Therefore, only a single sample PDS can be calculated over the entire observation
duration. In these cases the statistical distribution of the PDS remains to be
determined in the more general case of red deterministic variability due to the source.
One has then to be careful with assuming the 100\% uncertainties on the PDS
provided by standard tools such as {\tt powspec}: indeed, this relies on the
$\chi^2_2$ distribution, which does not hold any more in the more general case
of a variable source with red noise.

In the GRB literature, there have been different approaches. In one case, if
one considers a set of different time series due to {\em different} GRBs as
many realisations of the same stochastic process, then averaging the PDS of
different GRBs still make sense. However, it must be pointed out that a strong
assumption lies behind this: i.e., there is a unique stochastic process
giving rise to the variety of observed GRB time profiles.
This way one gains insight into the properties of this general process,
whose PDS is found to be described with a power--law, PDS $\propto f^{-5/3}$
\citep{Beloborodov00,Spada00}.
In the other case, each GRB time profile is considered individually as the
unique sample of a unique stochastic process, which is different from other
GRBs. The statistics of its unique PDS is not known and is no more a $\chi^2_2$.
Monte Carlo simulations aimed at estimating the PDS uncertainties by
generating other (synthetic) samples of the same process cannot make use of
the observed sample curve (e.g., \citealt{Ukwatta11}). Indeed, such a
procedure increases the noise variance and changes the $\chi^2_2$ nature
itself of the power distribution.
% Taking a Poisson process with expected value $\mu$
% and iteratively assuming its counts as expected counts of another Poisson
% process, after $n$ iterations, one ends up with a process which is no more
% Poisson, and whose expected value and variance are $\mu$ and $n\,\mu$,
% respectively \citep{Borgonovo07}. As a consequence, the distribution of the
% power is significantly changed.

In this work we address the issue of a correct evaluation of the statistics
of the PDS, and in particular of the power uncertainties, for a single sample
of a nonstationary and short--lived signal, such as that of GRB time profiles.
In this work time series are meant to be deterministic profiles affected
by uncorrelated noise.
We derived a formula for the variance of the PDS and show that it agrees with
the known results of L83 in the pure white noise case.
Finally, we test the validity of our results in the case of synthetic
light curves of typical GRBs.

Hereafter, time series are assumed to be discrete, equispaced, and with no data
gaps. Our treatment assumes the noise to be purely statistical
and uncorrelated, and we consider the two cases of Poisson or Gaussian
statistics, suitable to photon counting detectors.
PDSs are calculated assuming the Leahy normalisation (L83).
We therefore neglect dead time effects, which are known to affect the
statistics and suppress the variance of the resulting time series
(e.g., \citealt{Muller73}, K88).

%%%%%%%%%%%%%%%%%%%%%%%%%%%%%%%%%%%%%%%%%%%%%%
\section{Description}
\label{sec:desc}
%%%%%%%%%%%%%%%%%%%%%%%%%%%%%%%%%%%%%%%%%%%%%%
Let $x_k$ ($k=0,\ldots,N-1$) be a time series observed within
a time window with a duration $T$. The corresponding bin times
are $t_k=k\,T/N$. The observed series represents a sample function
of the true time series we would have observed in the absence
of statistical noise (e.g., with an infinite collecting area detector).
Hereafter, random variables are written in bold.
Each $x_k$ is therefore a single sample of the random variable
$\bmath{x}_k$, whose expected value and variance are defined
as $E\{\bmath{x}_k\}=\eta_k$, and
$E\{(\bmath{x}_k-\eta_k)^2\}=\sigma^2_k$.
The random variables are assumed to be independent,
$E\{\bmath{x}_k\,\bmath{x}_l\}=\eta_k\,\eta_l$ ($k\ne\,l$).

The discrete Fourier transform (DFT) amplitudes are defined by
\begin{eqnarray}
\bmath{a}_j & = & \sum_{k=0}^{N-1} \bmath{x}_k\,e^{2 \pi i j k/N} \qquad\qquad j=-\frac{N}{2}+1,\ldots, \frac{N}{2}\\
\bmath{x}_k & = & \frac{1}{N}\ \sum_{j=-N/2}^{N/2-1} \bmath{a}_j\,e^{-2 \pi i j k/N} \qquad k=0,\ldots, N-1 ,
\end{eqnarray}
where the corresponding $j$-th frequency is $f_j=j/T$.
The highest frequency is the Nyquist frequency, $f_{N/2}=N/2\,T$.
In the Leahy normalisation the power spectrum is defined by
\begin{equation}
\bmath{P}_j\ =\ \frac{2}{N_{\rm ph}}\ |\bmath{a}_j|^2\ =\ \frac{2}{N_{\rm ph}}\ 
\sum_{k,l} \bmath{x}_k\,\bmath{x}_l\,e^{2 \pi i (k-l) j/N}\qquad\Big(j=1,\ldots,\frac{N}{2}\Big)
\label{eq:p1}
\end{equation}
or, equivalently,
\begin{equation}
\bmath{P}_j\ =\ \frac{2}{N_{\rm ph}}\ \sum_{k,l} \bmath{x}_k\,\bmath{x}_l\,\cos{\Big(2 \pi (k-l) j/N\Big)} .
\label{eq:p2}
\end{equation}
$N_{\rm ph}=\sum_{k=0}^{N-1} \sigma^2_k$, is the expected total variance.
As shown in Section~\ref{sec:app_B}, this happens to coincide with the total number of
counts in the specific case of Poisson statistics, which explains the reason for the
choice of this name.
From equation~(\ref{eq:p1}) the expected value of the random variable $\bmath{P}_j$
is derived straightaway, and is
\begin{equation}
E\{\bmath{P}_j\}\ =\ 2 + \frac{2}{N_{\rm ph}}\ \sum_{k,l} \eta_k\,\eta_l\ e^{2 \pi i (k-l) j/N}\quad ,
\label{eq:exp_p}
\end{equation}
where we used $E\{\bmath{x}_k^2\}=\sigma^2_k + \eta^2_k$. The first term in the right-hand side
of equation~(\ref{eq:exp_p}) accounts for the statistical noise variance, also called ``white''
because it does not depend on frequency.
When the signal is constant, $\eta_k=\eta$ ($\forall k$), $\bmath{P}_j$ $(j=1,\ldots,N/2-1)$ is known
to be $\chi^2_2$ distributed, except for the Nyquist frequency, for which $P_{N/2}/2$ is
$\chi^2_1$ distributed and must be treated separately (L83).
In the pure white noise case the second term vanishes, thus leaving $E\{\bmath{P}_j\}=2$.

In general, the second term can also be seen as the DFT of the autocorrelation function (ACF)
defined by
\begin{equation}
R_j\ =\ \sum_{k=0}^{N-1} \eta_k\,\eta_{k+j}\qquad ,\quad (j=0, \ldots, N-1),
\label{eq:acf}
\end{equation}
assuming the periodic boundary condition $\eta_{k+N}=\eta_k$.
This is the discrete version of the Wiener--Khinchin theorem stating that the power
spectrum of a time series is the Fourier transform of its ACF (e.g., \citealt{Papoulis02}).

We define the power density spectrum of the deterministic function $P^{(\eta)}_j$ by
\begin{equation}
P^{(\eta)}_j\ =\ \frac{2}{N_{\rm ph}}\ \Big|\sum_{k=0}^{N-1} \eta_k e^{2 \pi i k j/N} \Big|^2
\label{eq:p_det}
\end{equation}
so that equation~(\ref{eq:exp_p}) can be written as
\begin{equation}
E\{\bmath{P}_j\}\ =\ 2 + P^{(\eta)}_j
\label{eq:exp_p2}
\end{equation}

So far no assumption was made on the kind of the distribution of the random variables
$\bmath{x}_k$. Hereafter, we distinguish between the Gaussian and the Poisson noise cases.

%------------------------------------
\subsection{Gaussian noise}
\label{sec:gaussian}
%------------------------------------
Each random variable $\bmath{x}_k$ is distributed according to a normal $N(\eta_k,\sigma_k)$.
We can express equation~(\ref{eq:p2}) as the result of matrix products:
\begin{equation}
\bmath{P}_j\ =\ \bmath{X}^T\, \bmath{A}\ \bmath{X}\quad , \quad
A_{kl}\ =\ \frac{2}{N_{\rm ph}}\,\cos{\Big(2 \pi (k-l) j/N\Big)}
\label{eq:p3}
\end{equation}
where $\bmath{X}$ is the column vector whose $k$-th row element is $\bmath{x}_k$, and
$A_{kl}$ is the $(k,l)$ element of $\bmath{A}$. The matrix $\bmath{A}$ is a positive-definite
quadratic form.
As such, from the algebra of the quadratic forms (e.g., \citealt{Ennis93}) the distribution
of the random variable $\bmath{P}_j$ is a non-central $\chi^2_r(\lambda)$ if and only if the
two following conditions are fulfilled:
\begin{eqnarray}
\displaystyle \left\{\begin{array}{lcl}
\displaystyle \bmath{A} & = & \bmath{A}\ \bmath{\Sigma}\ \bmath{A}\label{eq:c1}\\
\textrm{rank}(\bmath{A}) & = & r\label{eq:c2}\\
\end{array}
\right. \quad ,
\end{eqnarray}
where $r$ is the degrees of freedom, and
\begin{equation}
\lambda=\bmath{H}^T\bmath{A}\,\bmath{H}\ =\ P_j^{(\eta)}\quad ,
\label{eq:lambda}
\end{equation}
with $\bmath{H}$ being the column vector whose $k$-th element is $\eta_k$.
Equation~(\ref{eq:lambda}) follows from the definition of $P_j^{(\eta)}$ given
in equation~(\ref{eq:p_det}).
$\Sigma$ is the covariance matrix, so its $(k,l)$ element is given by
\begin{equation}
\Sigma_{kl}\ =\ \textrm{Cov}(\bmath{x}_k,\bmath{x}_l)\ =\ E\{\bmath{x}_k\,\bmath{x}_l\} -
E\{\bmath{x}_k\}\,E\{\bmath{x}_l\}\ =\ \delta_{kl}\,\sigma^2_k ,
\end{equation}
where $\delta_{kl}$ is Kronecker's delta.
To verify the first equation~(\ref{eq:c1}), we calculate the $(k,l)$ element of
the matrix $\bmath{A}\,\bmath{\Sigma}\,\bmath{A}$, which is found to be:
\begin{equation}
(\bmath{A}\,\bmath{\Sigma}\,\bmath{A})_{kl}\ =\ A_{kl} + \frac{2}{N_{\rm ph}^2}
\sum_{m=0}^{N-1} \sigma^2_m\,\cos{\Big(2 \pi (k+l-2m) j/N\Big)}\qquad\qquad\Big(j=1,\ldots,\frac{N}{2}-1\Big)
\label{eq:c1_test}
\end{equation}
The second term of the right-hand side is identically zero when all the variances
$\sigma^2_k$ are equal.
More generally, the second term of the right-hand side of equation~(\ref{eq:c1_test})
is O$(1/N_{\rm ph})$ times $A_{kl}$. Therefore, equation~(\ref{eq:c1}) is
approximately fulfilled in practical cases of interest.

The rank $r$ of $\bmath{A}$ does not depend on $\bmath{H}$, so we
can calculate it in the special case when $\eta_k=\eta$ ($\forall \eta$), i.e. the
case of constant signal. We already know that $\bmath{P}_j$ is $\chi^2_2$
distributed (L83), so it must be $r=2$.

This proves that in the more general case of a varying signal described by a
deterministic function whose discrete values are given by $\eta_k$ ($k=0,\ldots,N-1$),
and affected by pure uncorrelated Gaussian noise $\sigma_k$, the
Leahy-normalised power spectrum $\bmath{P}_j$ distributes according to a non-central
$\chi^2_2(\lambda)$, where $\lambda$ is $P_j^{(\eta)}$ (equation~\ref{eq:lambda}).
The expected value of $\bmath{P}_j$ is already known from equation~(\ref{eq:exp_p}),
while its variance is\footnote{If $\bmath{y}$ is $\chi^2_r(\lambda)$ distributed, then
$E\{\bmath{y}\}=r+\lambda$, and Var$(\bmath{y})=2\,(r+2\,\lambda)$.}
\begin{equation}
\textrm{Var}(\bmath{P}_j)\ =\ 2\,(2 + 2\,\lambda)\ =\
4\ \Big(1 + \frac{2}{N_{\rm ph}}\,\sum_{k,l}\,\eta_k\,\eta_l\,e^{2 \pi i (k-l) j/N}\Big)\ =\ 
4\ \Big(1 + P^{(\eta)}_j\Big)\qquad\qquad \Big(j=1,\ldots,\frac{N}{2}-1\Big)
\label{eq:var_p}
\end{equation}
Appendix~\ref{sec:app_A} reports the direct calculation of Var$(\bmath{P}_j)$ and
equation~(\ref{eq:A7}) reports the exact formula for the variance.
In the $j=N/2$ case $P_{N/2}/2$ satisfies the conditions~(\ref{eq:c1}) with $r=1$,
\begin{equation}
\textrm{Var}(\bmath{P}_{N/2})\ =\ 4\,\textrm{Var}(\bmath{P}_{N/2}/2)\ =\ 8\,(1 + 2\,\lambda)\ =\
8\ \Big(1 + \frac{2}{N_{\rm ph}}\,\sum_{k,l}\,\eta_k\,\eta_l\,e^{\pi i (k-l)}\Big)\ =\ 
8\ \Big(1 + P^{(\eta)}_{N/2}\Big)\quad ,
\label{eq:var_pNy}
\end{equation}
where we used $\lambda=P^{(\eta)}_{N/2}/2$.
Equations~(\ref{eq:var_p},\ref{eq:var_pNy}) can also be written as a function of the expected power:
\begin{eqnarray}
\displaystyle \textrm{Var}(\bmath{P}_j)\ =\ 
\displaystyle \left\{\begin{array}{l}
4\ (E\{\bmath{P}_j\} - 1)\qquad\qquad(j=1,\ldots,\frac{N}{2}-1)\\
8\ (E\{\bmath{P}_{N/2}\} - 1)\qquad\qquad(j=\frac{N}{2})
\end{array}
\right. \quad ,
\label{eq:var_p2}
\end{eqnarray}
Clearly, in the general case of uncorrelated noise and a significant power above
the white noise level, i.e. when $E\{\bmath{P}_j\}>2$, assuming a 100\% uncertainty
on the resulting power spectrum $\bmath{P}_j$, as assumed by widespread tools in X--ray astronomy
such as XRONOS {\tt powspec} overestimates the uncertainty. While the 100\% uncertainty
inherited from the $\chi^2_2$ distribution is correct under the assumption that all power is
due to Poisson noise and intrinsic correlated noise, in the case of uncorrelated noise here
considered from equation~(\ref{eq:var_p2}) the uncertainty is given by $2\,\sqrt{E\{\bmath{P}_j\}-1}$,
and not $E\{\bmath{P}_j\}$. The two coincide
only in the case of pure noise associated with a constant signal, so that $E\{\bmath{P}_j\}=2$.
The 100\% uncertainty assumed by XRONOS remains correct in the presence of genuine
stochastic variability due to correlated noise.

In practice, when only a single sample series $x_k$ is available of a short--lived process,
such as the time profile of a GRB, the deterministic series $\eta_k$ is not known a priori.
However, one could constrain the probability density function (pdf) of $\bmath{P}_j$, i.e. the
unknown $\lambda$ (equation~\ref{eq:lambda}), with the only sampled value $P_j$.
In principle, this makes no difference to taking
the light curve with observed $x_k$ counts as $x_k\pm\sqrt{x_k}$ instead of the unknown
$\eta_k\pm\sqrt{\eta_k}$ for a Poisson process (the Gaussian case is formally the same).
We estimate $\overline{\lambda}$, the best value for $\lambda$, adopting a Bayesian
approach:
\begin{equation}
p(r,\lambda | P_j)\ =\ \frac{p(P_j|r,\lambda)\,p(r,\lambda)}{p(P_j)}\quad ,
\end{equation}

where $p(r,\lambda | P_j)$ is the posterior function of the parameters $(r,\lambda)$ ($r$ is
fixed to either 1 or 2 depending on whether it is or not $j=N/2$) given the observed $P_j$.
The likelihood function $p(P_j|r,\lambda)$ is merely the pdf
of $\bmath{P_j}$, i.e. $\chi^2_r(\lambda, P_j)$. The prior $p(r,\lambda)$ may include the
knowledge one might have on $\lambda$ prior to measuring $P_j$, e.g. when a specific
shape of the deterministic PDS is expected. In the most general case, we assume a uniform
prior. The term $p(P_j)$ normalises the likelihood function.
For a given $P_j$ $\overline{\lambda}$ is chosen so as to maximise the posterior probability.
In this case our approach is equivalent to a maximum likelihood estimation (MLE).
In Appendix~\ref{sec:app_C} we show that it is $\overline{\lambda}=0$ ($P_j<2$), and
that $\overline{\lambda}$ rapidly converges to $P_j$ ($j<N/2$) and to $P_{N/2}/2$ ($j=N/2$)
for $P_j>2$. We conservatively assumed $\overline{\lambda}=P_j$ ($=P_{N/2}/2$ for $j=N/2$)
for all values of $P_j$, to avoid the risk of underestimating the variance.
Therefore, for a single time series $x_k$ consisting of an unknown deterministic function
affected by uncorrelated white noise equations~(\ref{eq:var_p}) and (\ref{eq:var_pNy}) are
approximated to
\begin{eqnarray}
\displaystyle \sigma(P_j) = \left\{\begin{array}{lr}
%\displaystyle 2 & \qquad (P_j\le 2, j<N/2)\\
\displaystyle 2\,\sqrt{P_j + 1} & \qquad (j<N/2)\\
%\displaystyle 2\,\sqrt{2} & \qquad (P_{N/2}\le 2)\\
\displaystyle 2\,\sqrt{2}\,\sqrt{P_{N/2} + 1} & \qquad (j=N/2)\\
\end{array}
\right. \quad .
\label{eq:var_p3}
\end{eqnarray}

Interestingly, the case of a constant variance (all $\sigma_k^2$ are equal), for which
the first equation~(\ref{eq:c2}) is fulfilled exactly (see equation~\ref{eq:c1_test}),
was discussed by \citet{Groth75}.
Apart from a scale factor of 2 in the definition of power, the probability density function
he derived is precisely that of a non-central chi square with $2\,n$ degrees of freedom,
and a non-central parameter given by the deterministic (or ``signal'', as he called it) power
(see equations~12 and 14 therein). His treatment considered the case where the power is
the sum of $n$ terms due to as many frequency bins, while here we consider the $n=1$ case.

%------------------------------------
\subsection{Poisson noise}
\label{sec:poisson}
%------------------------------------
Each random variable $\bmath{x}_k$ is distributed according to a Poisson distribution
with expected value $\eta_k$. Since $\bmath{x}_k$ are no more normal, we cannot exploit
the properties expressed by the conditions~(\ref{eq:c1}) for the Gaussian case.
In Appendix~\ref{sec:app_B} we calculate the corresponding variance of $\bmath{P}_j$.
Equation~(\ref{eq:B4}) reports the exact formula for Var$(\bmath{P}_j)$.

The main results are the following:
\begin{itemize}
\item if the observed counts $x_k$ are so high as to ensure the Gaussian regime ($x_k\gg 1$)
the results are the same as those discussed in Section~\ref{sec:gaussian}, since the overall
process essentially becomes Gaussian.
\item If $N_{\rm ph}\gg 1$, even if the individual $x_k$ are in the low-count regime, 
equations~(\ref{eq:var_p}--\ref{eq:var_p3}) still hold, and in particular $\bmath{P}_j$
still distributes according to a non-central $\chi^2_2(\lambda)$, as in Section~\ref{sec:gaussian}.
\item If $N_{\rm ph}\sim$~few, equations~(\ref{eq:var_p}--\ref{eq:var_p3}) do not hold any more
and the distribution of $\bmath{P}_j$ deviates from a $\chi^2_2(\lambda)$.
\end{itemize}

%%%%%%%%%%%%%%%%%%%%%%%%%%%%%%%%%%%%%
\section{Applications: gamma--ray burst light curves}
\label{sec:res}
%%%%%%%%%%%%%%%%%%%%%%%%%%%%%%%%%%%%%
We produced a number of GRB synthetic light curves to test the validity limits of the
results we derived for nature of the the power spectrum distribution and summarised by
equations~(\ref{eq:var_p}--\ref{eq:var_p3}). GRBs are particularly
suitable to this aim, given their nature of highly nonstationary and short--lived phenomena.
We adopted the fast-rise exponential decay (FRED) profile as modelled by \citet{Norris96}.
This function satisfactorily describes the temporal behaviour of the simplest example of
GRB light curve, which consists of a single pulse modelled as
\begin{eqnarray}
\displaystyle F(t) =
\displaystyle \left\{\begin{array}{l}
\displaystyle A\ \exp{\Big[-\Big(\frac{t_{\rm max}-t}{\tau_{\rm r}}\Big)^p\Big]}\quad,\quad t<t_{\rm max}\\
\displaystyle A\ \exp{\Big[-\Big(\frac{t-t_{\rm max}}{\tau_{\rm d}}\Big)^p\Big]}\quad,\quad t>t_{\rm max}\\
\end{array}
\right. \quad ,
\label{eq:norris}
\end{eqnarray}
where $t_{\rm max}$ is the peak time, $\tau_{\rm r}$ and $\tau_{\rm d}$ are the rise and
decay times, respectively, $A$ is the normalisation and $p$ is the peakedness (when $\nu=1$
the profile is a simple exponential, when $\nu=2$ it is a Gaussian).
For a typical FRED it is $\tau_{\rm r}/\tau_{\rm d}<1$, with an average value of $\sim0.3$--$0.5$
\citep{Norris96}.
The continuous power density spectrum of a FRED with $p=1$ (double exponential) can be calculated
analytically, and apart from a normalisation term is found to be
\begin{equation}
P(\nu)\ =\ \Big| \int_{-\infty}^{+\infty} F_{p=1}(t)\ e^{2 \pi i \nu t}\ dt\ \Big|^2\ =\ 
\frac{A^2\ (\tau_{\rm r} + \tau_{\rm d})^2}{\big[1 + (2\,\pi\,\nu\,\tau_{\rm r})^2\big]\ 
\big[1 + (2\,\pi\,\nu\,\tau_{\rm d})^2\big]}
\label{eq:pow_norris_anal}
\end{equation}
Several examples of power spectra obtained for the more general case of a FRED with $p\ne 1$
are discussed by \citet{Lazzati02}.

%
% ++++++++++++++ FREDINO 1 ++++++++++++++ Fig 1
\begin{figure}
\centering
\includegraphics[width=8.5cm]{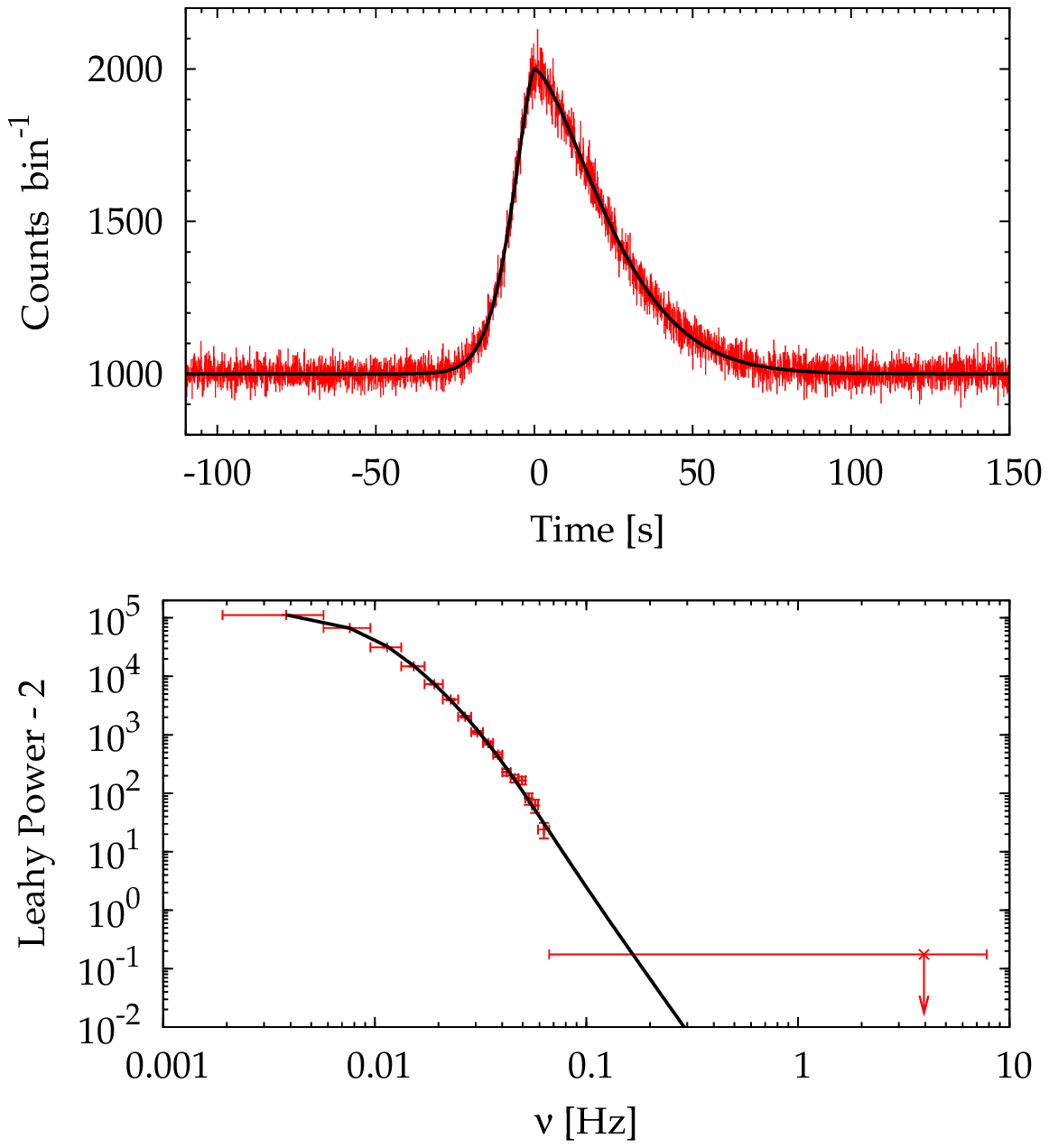}
\includegraphics[width=8.5cm]{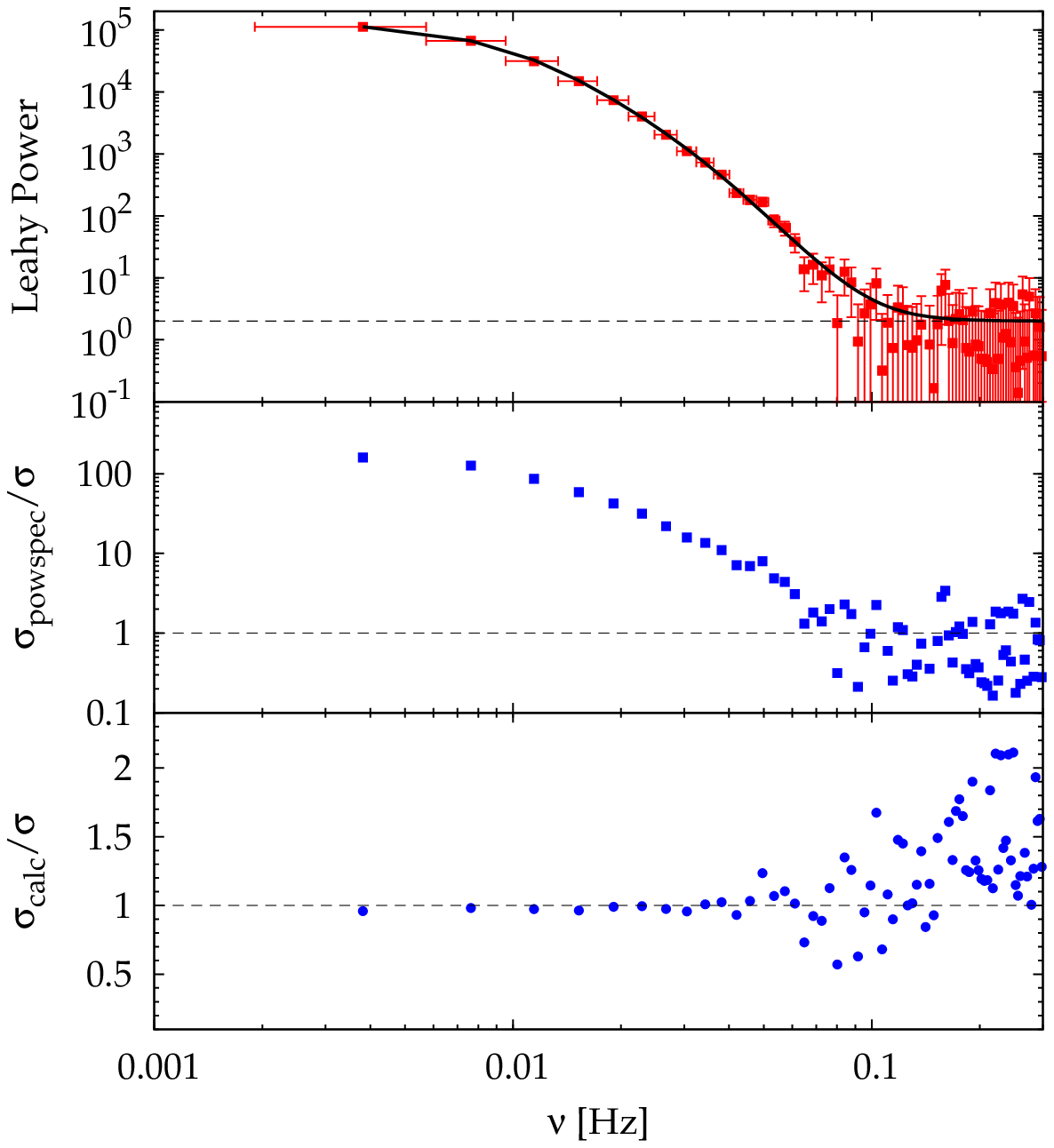}
\caption{{\em Top left}: example of a synthetic curve of a FRED-shaped GRB light curve. The thick solid
line represent the deterministic process.
{\em Top right}: power spectrum of the sample curve shown in the top left panel.
Uncertainties have been calculated following equation~(\ref{eq:var_p3}).
The solid line shows the power spectrum of the deterministic function, while the dashed line shows
the white noise level, 2.
{\em Mid right}: ratio between the uncertainties provided by the tool {\tt powspec} and the correct
values determined from the MC simulations for the corresponding frequency bin. 
{\em Bottom right}: ratio between the calculated uncertainties and the correct value from MC simulations.
{\em Bottom left}: the noise-subtracted power of the same sample function has been binned up so
as to ensure $3\sigma$ significance. The upper limit is at $3\sigma$.}
\label{f:fredino}
\end{figure}
%++++++++++++++++++++++++++++++++++++++++++
%
%
% ++++++++++++++ FREDINO power ++++++++++++++ Fig 1
\begin{figure}
\centering
\includegraphics[width=8.5cm]{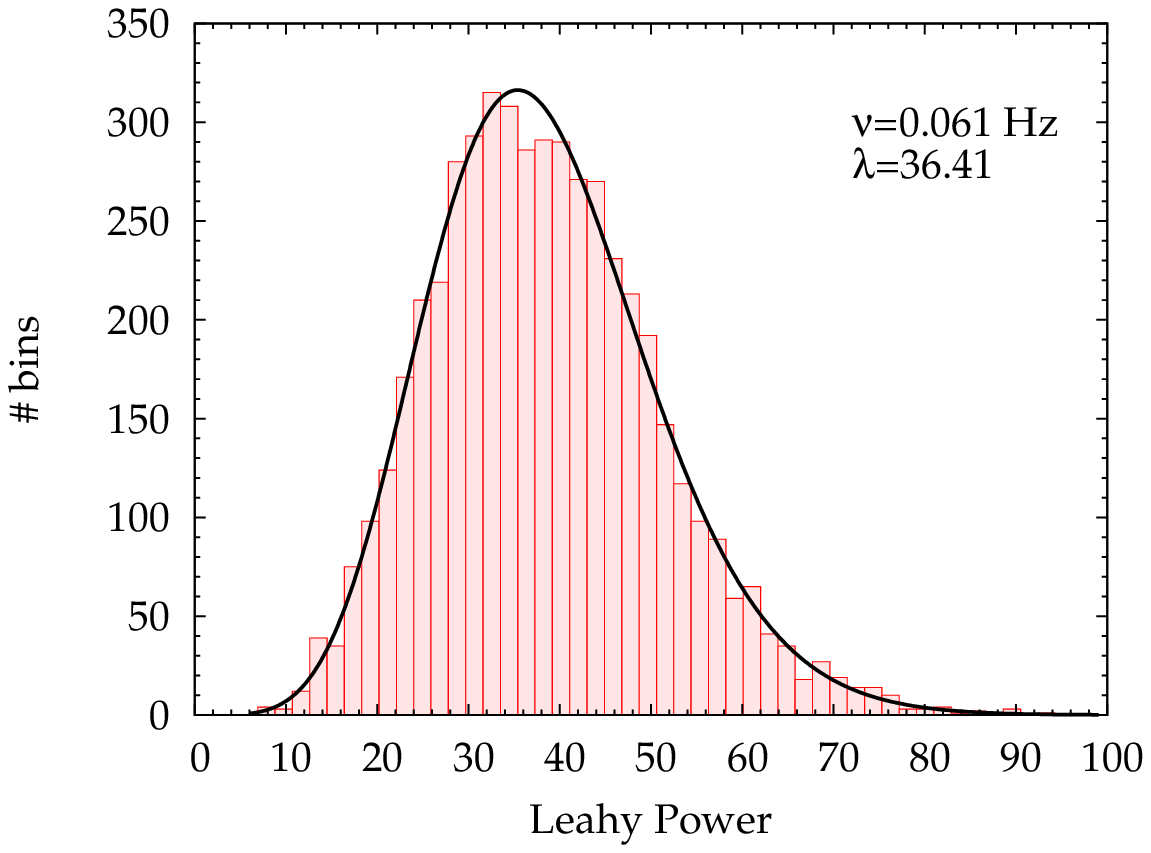}
\includegraphics[width=8.5cm]{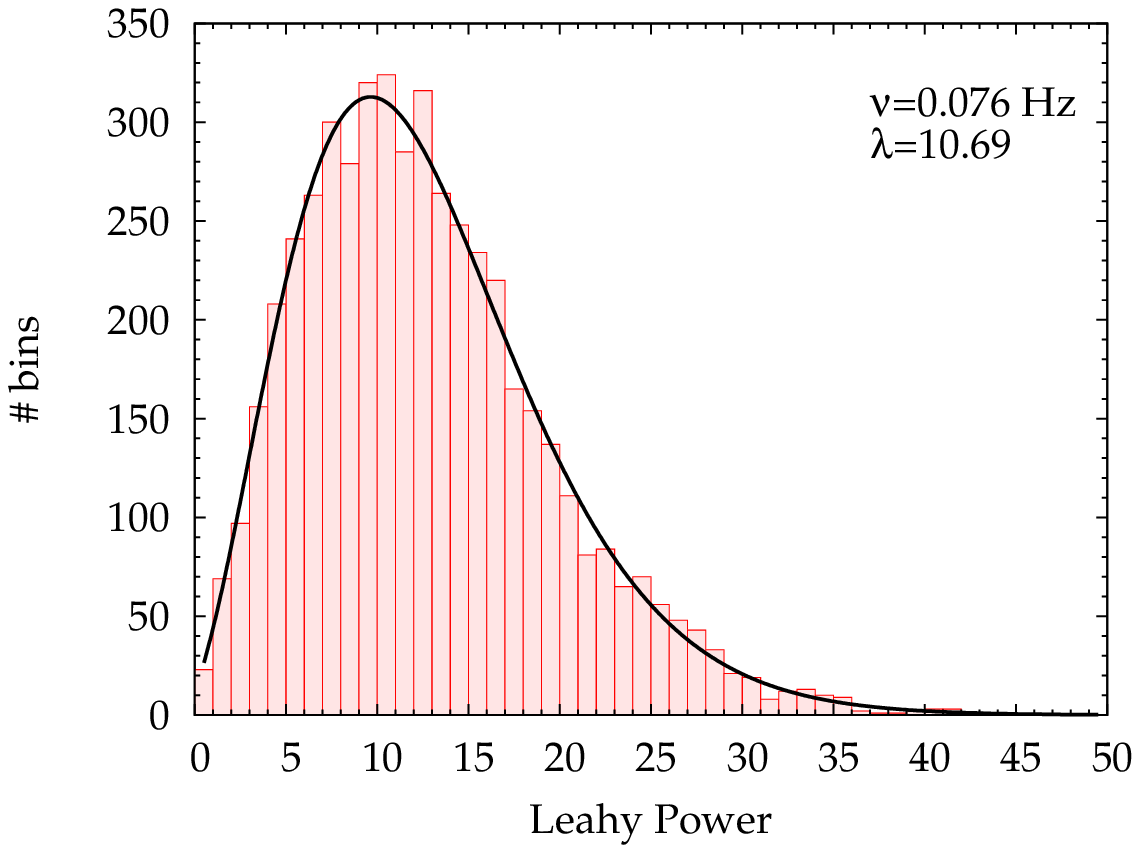}
\caption{{\em Left}: Leahy power distribution at $\nu=0.061$~Hz derived from 5000 synthetic sample
curves of the FRED pulse of Fig.~\ref{f:fredino}. The expected distribution is shown with a solid line,
and corresponds to a non-central $\chi^2_2(\lambda)$, with $\lambda=36.41$.
{\em Right}: the same at $\nu=0.076$~Hz, where $\lambda=10.69$.}
\label{f:distpow}
\end{figure}
%++++++++++++++++++++++++++++++++++++++++++
%
We started with a FRED with the following parameters: $\tau_{\rm r}=10$~s, $\tau_{\rm d}=30$~s,
$p=1.5$, $A=1000$~counts~bin$^{-1}$, $t_{\rm max}=0$~s, superposed to a constant detector background
with an average intensity of 1000~counts~bin$^{-1}$.
We generated $5\times10^3$ samples of this pulse with a bin time of $64$~ms, assuming Poisson statistics.
The total number of bins amounts to 4096 ($=2^{12}$).
Given the large number of counts per bin, this is equivalent to the Gaussian case.
Figure~\ref{f:fredino} displays the synthetic curve of the deterministic model as well as
one out of the simulated samples. The deterministic PDS is used for comparison to assess
the goodness of the PDS of the sample curve, in particular of the uncertainties derived
for the power adopting equation~(\ref{eq:var_p3}) as a function of frequency.
This is shown by the ratio between the calculated $\sigma$ and the scatter of the
power of the synthetic PDSs observed for each corresponding frequency bin
(bottom right panel of Fig.~\ref{f:fredino}). Apparently, the ratio ranges between $0.5$ and $2$.
An overall comparison with the analogous ratio between the uncertainty provided by {\tt powspec} and the
corresponding value determined from the MC simulations (mid right panel of Fig~\ref{f:fredino})
shows that the improvement is noteworthy.
Furthermore, the accuracy of equation~(\ref{eq:var_p3}) is evident when
the PDS is dominated by the deterministic power of the signal (at low frequencies).
At high frequencies, where the white statistical noise dominates, the power uncertainty
is systematically overestimated up to a factor of 2. However, compared to the values provided
by {\tt powspec} often underestimated by up to an order of magnitude, it still represents
a significant improvement, particularly in the more conservative direction of overestimating
rather than underestimating uncertainties.

So far this proves that equation~(\ref{eq:var_p3}) overall provides a satisfactory means for estimating
the uncertainty on the PDS of a sample curve. We go further and test whether the distribution
of the individual $\bmath{P}_j$ is indeed a non-central chi-square distribution. To this aim, from
the PDS of the sample curve considered above we chose two frequencies and derived the corresponding
power distribution from the synthetic PDSs. The result is displayed in Fig.~\ref{f:distpow}.
The expected non-central parameter for the corresponding $\chi^2_2(\lambda)$ was calculated
with equations~(\ref{eq:p_det}, \ref{eq:var_p}).

\begin{figure}
\centering
\includegraphics[width=8.5cm]{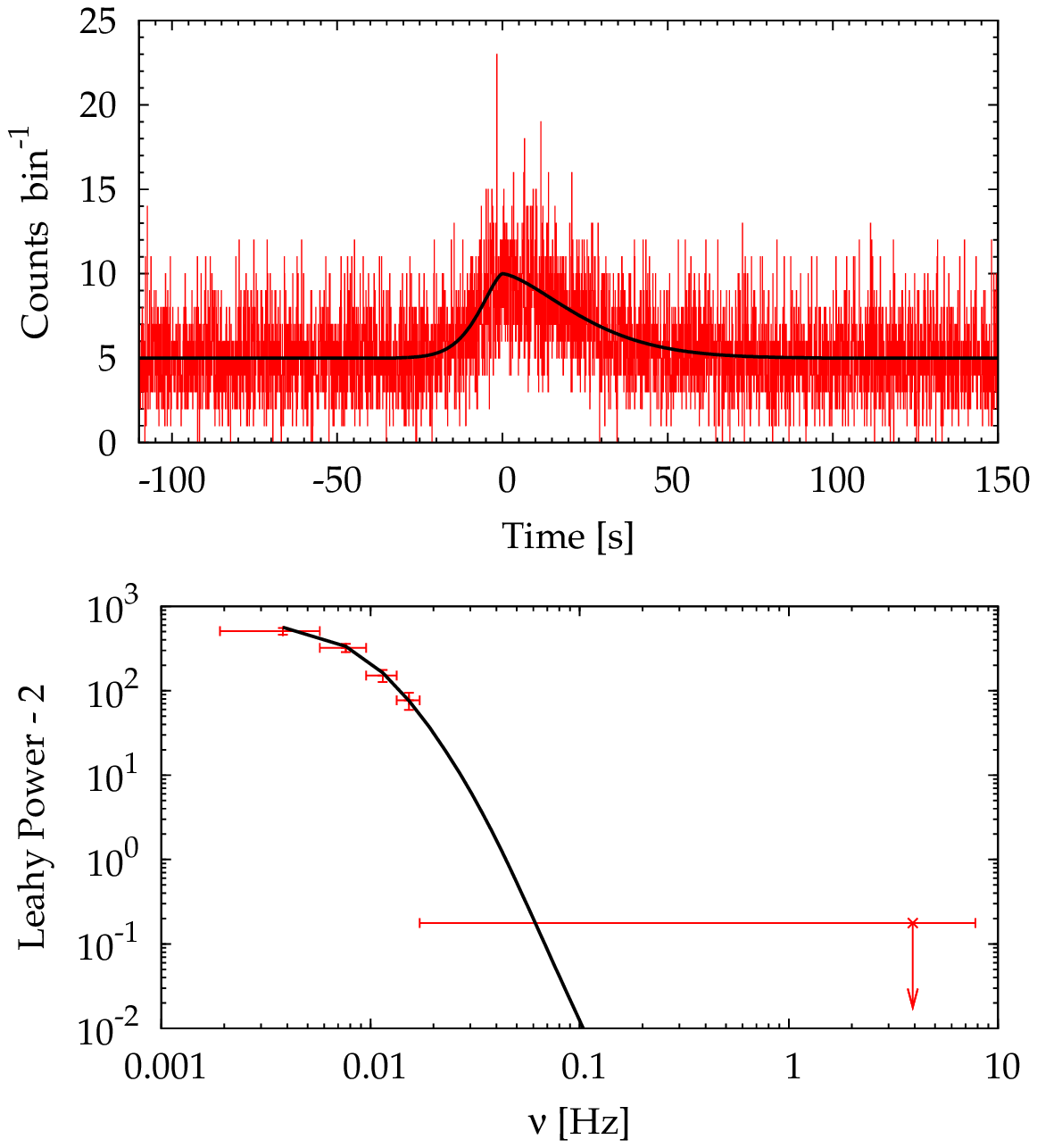}
\includegraphics[width=8.5cm]{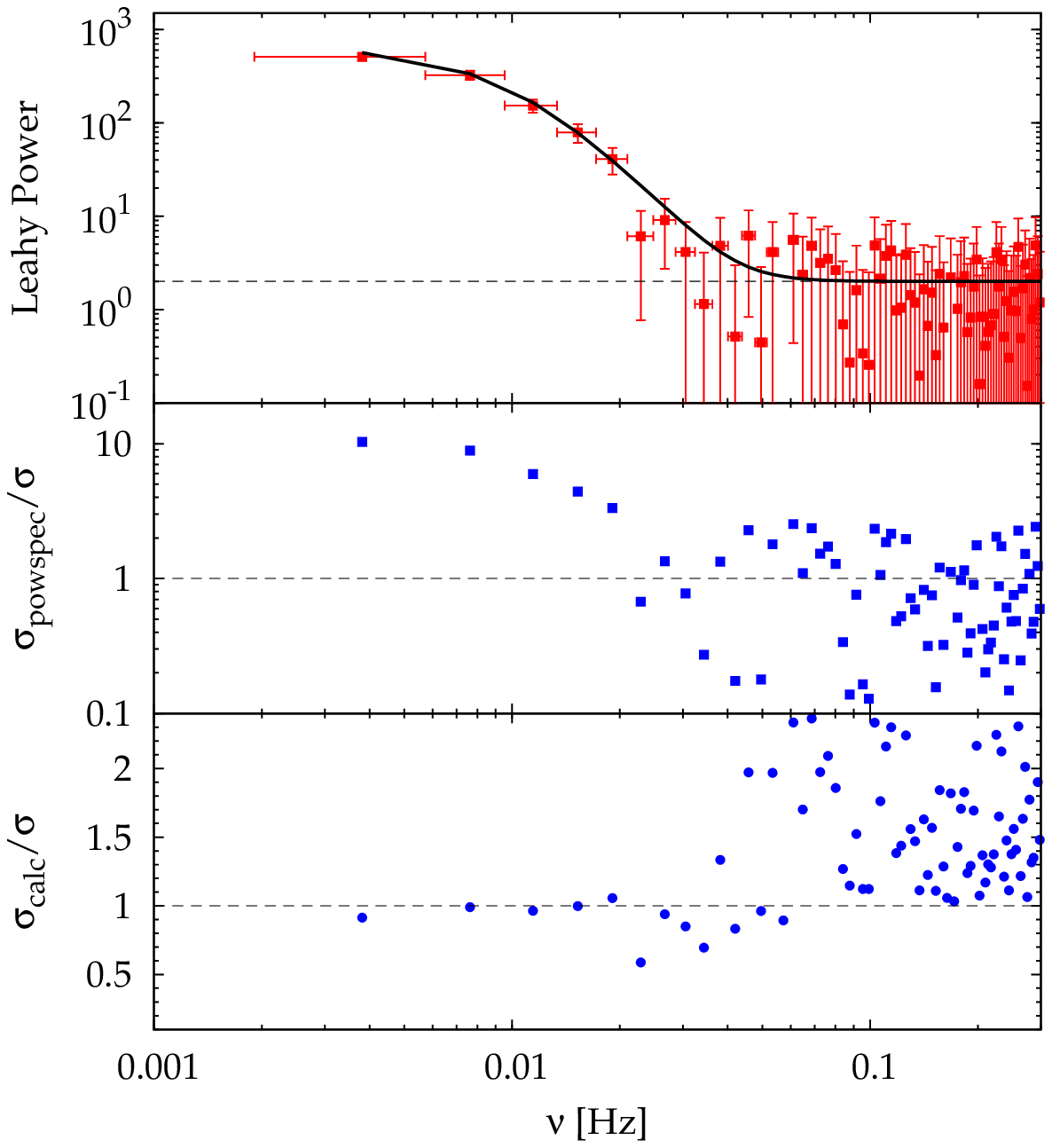}
\caption{Same as in Fig.~\ref{f:fredino}. Here the FRED has the same profile, but it is 200 times
less intense. The background level is proportionally lower, 5~counts~bin$^{-1}$. This example
fits in the low-count rate Poisson regime ($N_{\rm ph}=2.3\times10^4$).}
\label{f:minifredino}
\end{figure}

\begin{figure}
\centering
\includegraphics[width=8.5cm]{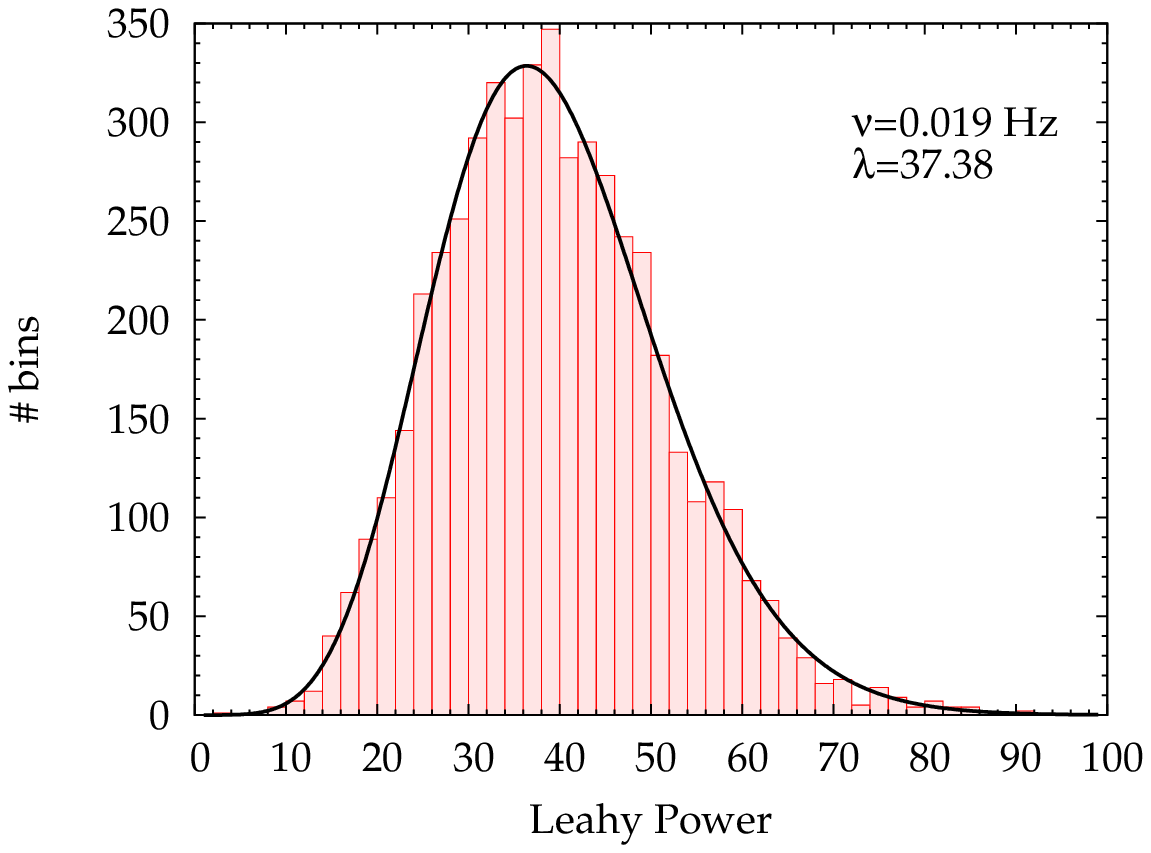}
\includegraphics[width=8.5cm]{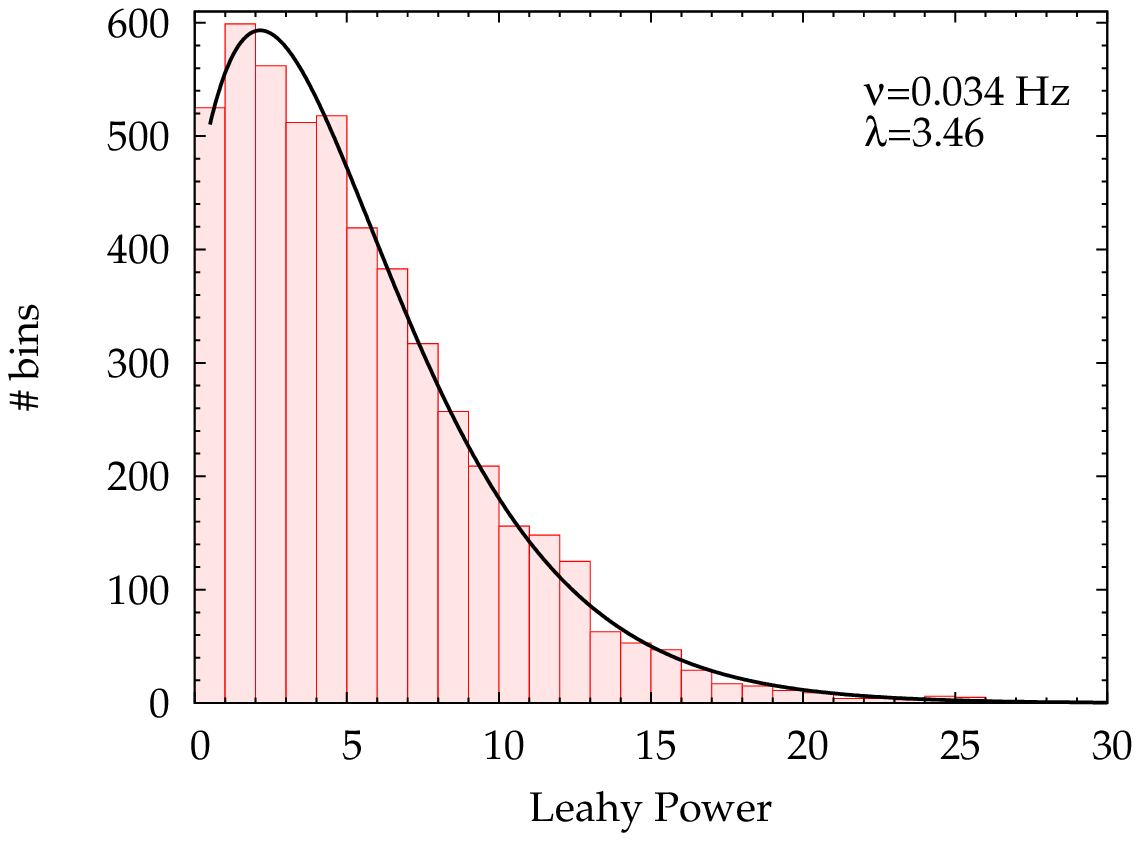}
\caption{Same as in Fig.~\ref{f:distpow}, referred to the pulse of Fig.~\ref{f:minifredino}.}
\label{f:distpow_mini}
\end{figure}

Figure~\ref{f:minifredino} shows the same FRED pulse 200 times fainter superposed to a correspondingly
fainter constant background of 5~counts~bin$^{-1}$. The single $\bmath{x}_k$ variables cannot be
approximately assumed to be normally distributed, so that we may test the Poisson regime.
Still, the total number of counts is still very large, $N_{\rm ph}=2.3\times10^4$. This means that
the results obtained for the Gaussian case should still hold.
Indeed, both the comparison between the calculated uncertainties and the scatter of the corresponding
power from the simulated PDSs gives similar results to the previous case (bottom right panel of
Fig.~\ref{f:minifredino}). Likewise, the power distributions of individual frequency bins are fully
compatible with the corresponding expected non-central chi-squares, as shown by Fig.~\ref{f:distpow_mini}.
%
% ++++++++++++++ MINIFREDINO 1 ++++++++++++++ Fig 1
%++++++++++++++++++++++++++++++++++++++++++
%

%
% ++++++++++++++ MINIFREDINO power ++++++++++++++ Fig 1
%++++++++++++++++++++++++++++++++++++++++++
%

%%%%%%%%%%%%%%%%%%%%%%%%%%%%%%%%%%%%%
\section{Summary and Conclusions}
\label{sec:conc}
%%%%%%%%%%%%%%%%%%%%%%%%%%%%%%%%%%%%%
We investigated the nature of the statistical distribution of the power density spectrum of
a single, nonstationary, and short--lived time profile on a theoretical ground.
This treatment assumes the time series to be deterministic profiles affected by
uncorrelated noise. In other words,
the time series here considered consist of a set of statistically independent,
Poisson and normally distributed random variables, whose expected values represent
the deterministic function of the varying signal to be studied.

We demonstrated that the probability density function of the power is a non-central
$\chi^2_2(\lambda)$, whose non-central parameter $\lambda$ corresponds to the power of
the deterministic function. This holds in the Gaussian case, as well as in the Poisson
case, provided that the Gaussian limit of $N_{\rm ph}\gg 1$ is fulfilled ($N_{\rm ph}$
being the total number of counts).
As a consequence, we provided a new formula for calculating the correct uncertainty
of the power at each frequency as a function of the observed power itself.
We finally showed the agreement with simulated light curves of typical GRB time profiles.
These results provide a statistically solid basis to a proper treatment of power density
spectra in the case of nonstationary and short--lived time series affected by
uncorrelated noise.

\section*{Acknowledgments}

The author is grateful to Mauro Orlandini and Raffaella Margutti for reading the manuscript
and for their useful comments. The author also wishes to thank the referee Michiel
van~der~Klis for the very useful comments. The author acknowledges ASI for financial support
(ASI-INAF contract I/088/06/0).

\appendix

%%%%%%%%%%%%%%%%%%%%%%%%%%%%%%%%%
\section{Variance of the power (Gaussian case)}
\label{sec:app_A}
%%%%%%%%%%%%%%%%%%%%%%%%%%%%%%%%%
The central moments of a random variable $\bmath{x}_k$ normally distributed as
$N(\eta_k,\sigma_k)$ are
\begin{equation}
\label{eq:A1}
E\{(\bmath{x}_k-\eta_k)^{2 i + 1}\} = 0 \quad (\forall i),\qquad
E\{(\bmath{x}_k-\eta_k)^2\} = \sigma^2_k ,\qquad
E\{(\bmath{x}_k-\eta_k)^4\} = 3 \sigma^4_k
\end{equation}
where the different variables at different values of $k$ are meant to be independent,
so $E\{\bmath{x}_k\bmath{x}_l\} = \eta_k \eta_l$, $(k\ne l)$. They can be used to calculate the
corresponding  noncentral moments through the following,
\begin{equation}
\label{eq:A2}
E\{\bmath{x}^n_k\}\ = \ \sum_{i=0}^{n} {n \choose i} E\{(\bmath{x}_k-\eta_k)^i\}\ \eta^{n-i}_k .
\end{equation}
The noncentral moments are given by
\begin{equation}
E\{\bmath{x}_k\} = \eta_k ,\qquad
E\{\bmath{x}^2_k\} = \eta^2_k + \sigma^2_k ,\qquad
E\{\bmath{x}^3_k\} = \eta^3_k + 3 \eta_k \sigma^2_k ,\qquad
E\{\bmath{x}^4_k\} = \eta^4_k + 6 \eta^2_k \sigma^2_k + 3\sigma^4_k .
\label{eq:A3}
\end{equation}
The variance of $\bmath{P}_j$ is calculated directly from equation~(\ref{eq:p1}):
\begin{eqnarray}
\label{eq:A4}
\Big(\frac{N_{\rm ph}}{2}\Big)^2\,\textrm{Var}(\bmath{P}_j) & = &
\Big(\frac{N_{\rm ph}}{2}\Big)^2\,\Big(E\{\bmath{P}^2_j\}-E^2\{\bmath{P}_j\}\Big)\\\nonumber
& = & \sum_{k,l,m,n} E\{\bmath{x}_k\bmath{x}_l\bmath{x}_m\bmath{x}_n\} e^{2 \pi i (k-l+m-n) j/N} -
\Big(\sum_k \sigma^2_k + \sum_{k,l} \eta_k \eta_l  e^{2 \pi i (k-l) j/N}\Big)^2 \quad ,
\end{eqnarray}
where we used the definition of $N_{\rm ph}$. The terms of equation~(\ref{eq:A4}) must conveniently
be separated based on the different kind of moments. To simplify the notation, we define
$\omega=2\pi j/N$.
\begin{eqnarray}
\label{eq:A5}
\displaystyle
\Big(\frac{N_{\rm ph}}{2}\Big)^2\,\textrm{Var}(\bmath{P}_j)& = &
\sum_k E\{\bmath{x}^4_k\}
+ 2\,\sum_{k\ne l} E\{\bmath{x}^3_k\} E\{\bmath{x}_l\} \Big(e^{\omega i (k-l)} + e^{-\omega i (k-l)}\Big)
+ \sum_{k\ne l} E\{\bmath{x}^2_k\} E\{\bmath{x}^2_l\} \Big(2 + e^{2 \omega i (k-l)}\Big) +\\\nonumber
&& \sum_{k,l,m,\ne} E\{\bmath{x}^2_k\} E\{\bmath{x}_l\} E\{\bmath{x}_m\}
   \Big(4 e^{\omega i (l-m)} + e^{\omega i (2k-l-m)} +  e^{-\omega i (2k-l-m)}\Big) + \\\nonumber
&& \sum_{k,l,m,n,\ne} E\{\bmath{x}_k\} E\{\bmath{x}_l\} E\{\bmath{x}_m\} E\{\bmath{x}_n\}
   e^{\omega i (k-l+m-n)}
-\sum_{k,l} \sigma^2_k \sigma^2_l -\sum_{k,l,m,n} \eta_k \eta_l \eta_m \eta_n e^{\omega i (k-l+m-n)}\\\nonumber
&& -2\,\sum_{k,l,m} \sigma^2_m \eta_k \eta_l e^{\omega i (k-l)} .
\end{eqnarray}
We adopted the following notation: $\sum_{k,l,m,\ne}$ is a sum over $k,l$, and $m$, and is meant to
exclude all the equality cases ($k\ne l$, $k\ne m$, $l\ne m$).
Replacing the values of the corresponding moments,
\begin{eqnarray}
\label{eq:A6}
\displaystyle
& = &
\sum_k (6 \eta_k^2 \sigma_k^2 + 3 \sigma_k^4)
+ 6\,\sum_{k\ne l} \sigma_k^2 \eta_k \eta_l \Big(e^{\omega i (k-l)} + e^{-\omega i (k-l)}\Big)
+ \sum_{k\ne l} (\sigma^2_k\sigma_l^2 + \eta_k^2\sigma_l^2 + \eta_l^2\sigma_k^2)\,
\Big(2 + e^{2 \omega i (k-l)}\Big) +\\\nonumber
& & \sum_{k,l,m,\ne} \sigma_k^2 \eta_l \eta_m
   \Big(4 e^{\omega i (l-m)} + e^{\omega i (2k-l-m)} +  e^{-\omega i (2k-l-m)}\Big)
-\sum_{k,l} \sigma^2_k \sigma^2_l -2\,\sum_{k,l,m} \sigma^2_m \eta_k \eta_l e^{\omega i (k-l)}
\end{eqnarray}
After a few passages, by adding and subtracting the terms excluded in the sums
in equation~(\ref{eq:A6}), one ends up with the following:
\begin{eqnarray}
\label{eq:A7}
\displaystyle
\textrm{Var}(\bmath{P}_j)& = &
4\ \Big(1 + \frac{2}{N_{\rm ph}}\,\sum_{k,l} \eta_k \eta_l e^{2 \pi i (k-l) j/N}\ \Big) +\\\nonumber
& & \frac{4}{N_{\rm ph}^2}\, \Big[\sum_{k,l} \sigma^2_k \sigma_l^2 e^{4 \pi i (k-l) j/N} +
\sum_{k,l,m} \sigma^2_k \eta_l \eta_m (e^{2 \pi i (2k-l-m) j/N} + e^{-2 \pi i (2k-l-m) j/N})\ \Big] .
\end{eqnarray}
Equation~(\ref{eq:A7}) is exact. For the Nyquist frequency, equation~(\ref{eq:A7}) becomes
\begin{equation}
\label{eq:A7Ny}
\textrm{Var}(\bmath{P}_{N/2})\ =\ 
8\ \Big(1 + \frac{2}{N_{\rm ph}}\,\sum_{k,l} \eta_k \eta_l e^{\pi i (k-l)}\ \Big)
\end{equation}
in agreement with equation~(\ref{eq:var_pNy}).
For $j\ne N/2$ the second term in the right-hand side of equation~(\ref{eq:A7}) is identically
zero when all $\sigma_k=\sigma$ ($\forall k$), and it can be neglected
when the $\sigma_k$'s are comparable with each other, being O$(1/N_{\rm ph})$ times
the first term. Equation~(\ref{eq:A7}) can be approximated by
\begin{eqnarray}
\displaystyle \textrm{Var}(\bmath{P}_j)\ 
\displaystyle \left\{\begin{array}{l}
\ \simeq\ 4\ \Big(1 + \frac{2}{N_{\rm ph}}\,\sum_{k,l} \eta_k \eta_l
e^{2 \pi i (k-l) j/N}\ \Big)\qquad\qquad\Big(j=1,\ldots,\frac{N}{2}-1\Big)\\
\ =\ 8\ \Big(1 + \frac{2}{N_{\rm ph}}\,\sum_{k,l} \eta_k \eta_l
e^{\pi i (k-l)}\ \Big)\qquad\qquad\qquad\Big(j=\frac{N}{2}\Big)
\end{array}
\right. \quad ,
\label{eq:A8}
\end{eqnarray}
in agreement with equations~(\ref{eq:var_p},\ref{eq:var_pNy}).

%%%%%%%%%%%%%%%%%%%%%%%%%%%%%%%%%
\section{Variance of the power (Poisson case)}
\label{sec:app_B}
%%%%%%%%%%%%%%%%%%%%%%%%%%%%%%%%%
The case of a process $\bmath{x}_k$ affected by Poisson noise is a more general
case than that of a Gaussian noise, since the latter corresponds to the former
in the high count rate regime, i.e. when it is $E\{\bmath{x}_k\}=\eta_k\gg 1$.
For a Poisson process, the central moments are
\begin{equation}
\label{eq:B1}
E\{(\bmath{x}_k-\eta_k)^2\} = \eta_k ,\qquad
E\{(\bmath{x}_k-\eta_k)^3\} = \eta_k ,\qquad
E\{(\bmath{x}_k-\eta_k)^4\} = \eta_k + 3 \eta^2_k .
\end{equation}
From equation~(\ref{eq:A2}) the corresponding noncentral moments are
\begin{equation}
\label{eq:B2}
E\{\bmath{x}_k\} = \eta_k ,\qquad
E\{\bmath{x}_k^2\} = \eta_k^2 + \eta_k ,\qquad
E\{\bmath{x}_k^3\} = \eta_k^3 + 3 \eta_k^2 + \eta_k ,\qquad
E\{\bmath{x}_k^4\} = \eta_k^4 + 6 \eta_k^3 + 7 \eta_k^2 + 3 \eta_k .
\end{equation}
In the Poisson case the normalisation constant $N_{\rm ph}=\sum_{k=0}^{N-1} \eta_k$ is
the expected total counts.
The definition of the power spectrum $\bmath{P}_j$ is the same as that of
equation~(\ref{eq:p1}). Using the first two moments of equation~(\ref{eq:B2}),
calculating the expected value of $\bmath{P}_j$ is straightforward, and is found
to be the same as equation~(\ref{eq:exp_p}).

Calculating the variance of $\bmath{P}_j$ in the Poisson case is formally the same
as the Gaussian case up to equation~(\ref{eq:A5}), i.e., prior to substituting the
specific values of the moments. At this point the two cases must be treated separately.
Replacing the moments of equation~(\ref{eq:B2}) in (\ref{eq:A5}), and defining
$\omega=2\,\pi\,j/N$ as before, it becomes
\begin{eqnarray}
\label{eq:B3}
\displaystyle
\Big(\frac{N_{\rm ph}}{2}\Big)^2\,\textrm{Var}(\bmath{P}_j)& = &
\sum_k (6 \eta_k^3 + 7 \eta_k^2 + \eta_k)
+ \sum_{k\ne l} (6 \eta_k^2 + 2 \eta_k) \eta_l \Big(e^{\omega i (k-l)} + e^{-\omega i (k-l)}\Big)
+ \sum_{k\ne l} \Big[2 \eta_k \eta_l + 4 \eta_k^2 \eta_l +\\\nonumber
& & (\eta_k \eta_l + \eta_k^2 \eta_k + \eta_k \eta_l^2) e^{2 \omega i (k-l)}\Big] +
\sum_{k,l,m,\ne} \eta_k \eta_l \eta_m \Big(4 e^{\omega i (l-m)} + e^{\omega i (2k-l-m)} +
e^{-\omega i (2k-l-m)}\Big)\\\nonumber
& &-\sum_{k,l} \eta_k \eta_l -2\,\sum_{k,l,m} \eta_k \eta_l \eta_m e^{\omega i (l-m)} .
\end{eqnarray}
Similarly to what was done in section~\ref{sec:app_A}, adding and subtracting the excluded
terms in the sums, after a few passages one ends up with
\begin{eqnarray}
\label{eq:B4}
\displaystyle
\textrm{Var}(\bmath{P}_j)& = &
4\ \Big(1 + \frac{1}{N_{\rm ph}}\Big) + \frac{8}{N_{\rm ph}}\,\Big(1 + \frac{2}{N_{\rm ph}}\Big)
\,\sum_{k,l} \eta_k \eta_l e^{2 \pi i (k-l) j/N} +\\\nonumber
& & \frac{4}{N_{\rm ph}^2}\, \Big[\sum_{k,l} \eta_k \eta_l e^{4 \pi i (k-l) j/N} +
\sum_{k,l,m} \eta_k \eta_l \eta_m (e^{2 \pi i (2k-l-m) j/N} + e^{-2 \pi i (2k-l-m) j/N})\ \Big] .
\end{eqnarray}
As done for the Gaussian case, we have to treat the Nyquist frequency separately.
When $j=N/2$, equation~(\ref{eq:B4}) becomes
\begin{equation}
\label{eq:B4Ny}
\textrm{Var}(\bmath{P}_{N/2})\ =\ 
4\ \Big(2 + \frac{1}{N_{\rm ph}}\Big) + \frac{8}{N_{\rm ph}}\,\Big(2 + \frac{2}{N_{\rm ph}}\Big)
\,\sum_{k,l} \eta_k \eta_l e^{\pi i (k-l)}\quad ,
\end{equation}
which is equivalent to equation~(\ref{eq:A7Ny}) in the $N_{\rm ph}\gg 1$ limit.
In the special case of a constant signal $\eta_k=\eta$ ($\forall k$),
equation~(\ref{eq:B4}) reduces to
\begin{eqnarray}
\displaystyle \textrm{Var}(\bmath{P}_j)\ =\ 
\displaystyle \left\{\begin{array}{l}
4\ \Big(1 + \frac{1}{N_{\rm ph}}\Big)\qquad\qquad\Big(j=1,\ldots,\frac{N}{2}-1\Big)\\
4\ \Big(2 + \frac{1}{N_{\rm ph}}\Big)\qquad\qquad\Big(j=\frac{N}{2}\Big)
\end{array}
\right. \quad ,
\label{eq:B5}
\end{eqnarray}
in agreement with the results of L83. In the more general case of a nonstationary signal
$\eta_k$, in the limit $N_{\rm ph}\gg 1$, equation~(\ref{eq:B4}) can be approximated by
\begin{eqnarray}
\label{eq:B6}
\displaystyle
\textrm{Var}(\bmath{P}_j)& \simeq &
4\ \Big(1 + \frac{2}{N_{\rm ph}} \sum_{k,l} \eta_k \eta_l e^{2 \pi i (k-l) j/N}\Big) +\\\nonumber
& & \frac{4}{N_{\rm ph}^2}\, \Big[\sum_{k,l} \eta_k \eta_l e^{4 \pi i (k-l) j/N} +
\sum_{k,l,m} \eta_k \eta_l \eta_m (e^{2 \pi i (2k-l-m) j/N} + e^{-2 \pi i (2k-l-m) j/N})\ \Big] .
\end{eqnarray}
Not surprisingly, equation~(\ref{eq:B6}) is the same as (\ref{eq:A7}) upon replacing
$\sigma_k^2$ with $\eta_k$, as expected for a Poisson variable in the Gaussian limit.
As discussed in Appendix~\ref{sec:app_A}, the result in equation~(\ref{eq:B6}) can be
approximated by equation~(\ref{eq:A8}), provided that $N_{\rm ph}\gg 1$.
When the Gaussian limit is not satisfied, i.e., when $N_{\rm ph}$ is just a few,
we note that the relation between expected value and variance for a noncentral chi-square
distributed random variable with $r=2$ degrees of freedom $(j\ne N/2)$ is not fulfilled:
\begin{eqnarray}
\label{eq:B7}
\displaystyle
E\{\bmath{P}_j\} & = & 2 + \frac{2}{N_{\rm ph}}\ \sum_{k,l} \eta_k \eta_l e^{2 \pi i (k-l) j/N}\ =\
r + \lambda\\
\label{eq:B8}
\textrm{Var}(\bmath{P}_j)& = & 2\,\Big(2 + \frac{4}{N_{\rm ph}}\
\sum_{k,l} \eta_k \eta_l e^{2 \pi i (k-l) j/N} \Big) + \frac{4}{N_{\rm ph}}\ \Big(1 + \frac{4}{N_{\rm ph}}\
\sum_{k,l} \eta_k \eta_l e^{2 \pi i (k-l) j/N}\Big)\ +\\\nonumber
& & + \frac{4}{N_{\rm ph}^2}\, \Big[\sum_{k,l} \eta_k \eta_l e^{4 \pi i (k-l) j/N} +
\sum_{k,l,m} \eta_k \eta_l \eta_m (e^{2 \pi i (2k-l-m) j/N} + e^{-2 \pi i (2k-l-m) j/N})\ \Big]\\\nonumber
& & =\ (r + 2 \lambda) + \frac{2}{N_{\rm ph}}\,(r + 4 \lambda) + \frac{4}{N_{\rm ph}^2}\Big[\ldots\Big]
\ne 2\ (r + 2 \lambda) .
\end{eqnarray}

We conclude that for $N_{\rm ph}\sim$~few, the distribution of the power spectrum at
a given frequency is not a noncentral $\chi^2_2(\lambda)$, as found in the Gaussian limit,
and the expression for the variance to be used is given by equation~(\ref{eq:B8}).
In the same regime of $N_{\rm ph}\sim$~few, $P_{N/2}/2$ also deviates from a noncentral
$\chi^2_1(\lambda)$ distribution, given that equation~(\ref{eq:B4Ny}) does not fulfil
any more the corresponding relation between expected value and variance. The exact
formula for the variance of the Nyquist frequency power therefore remains
equation~(\ref{eq:B4Ny}).

%%%%%%%%%%%%%%%%%%%%%%%%%%%%%%%%%
\section{Maximum likelihood estimation}
\label{sec:app_C}
%%%%%%%%%%%%%%%%%%%%%%%%%%%%%%%%%
The $j<N/2$ and $j=N/2$ cases must be treated separately, given that the probability
density functions of the corresponding random variables $\bmath{P}_j$ are non-central
chi squares with $r=2$ and $r=1$ degrees of freedom, respectively.
The purpose is to find the value for the non-central parameter $\overline{\lambda}$
which maximises the likelihood function $\chi^2_r(\lambda,P_j)$ for a given measured
value $P_j$.

\begin{figure}
\centering
\includegraphics[width=8.5cm]{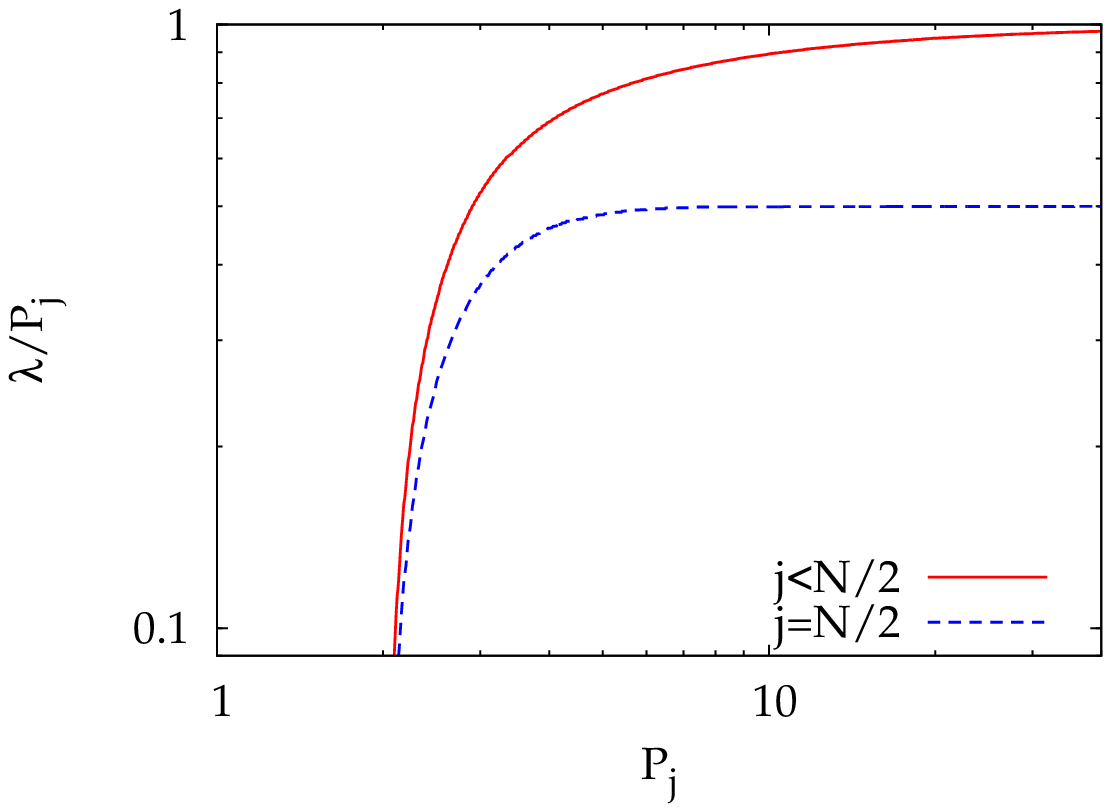}
\caption{Best estimate for the non-central $\chi^2$ distribution parameter $\overline{\lambda}$
as a function of $P_j$.}
\label{f:lambda}
\end{figure}

First, let us consider the $j<N/2$ case, for which it is $r=2$.
It is
\begin{equation}
\chi^2_2(\lambda, P_j)\ =\ \frac{1}{2\,\pi}\, e^{-(P_j + \lambda)/2}\ \int_0^\pi e^{\sqrt{\lambda\,P_j}\ \cos{\theta}}\ d\theta\quad .
\label{eq:chi2}
\end{equation}
When $P_j=0$ equation~(\ref{eq:chi2}) reduces to a simple exponential, so that
$\overline{\lambda}=0$. For $P_j\le 2$, it is $\overline{\lambda}=0$.
For $P_j>2$, $\overline{\lambda}$ is found by requiring
\begin{equation}
\frac{\partial \chi^2_2(\lambda, P_j)}{\partial \lambda}\Big|_{\lambda=\overline{\lambda}}\ =\ 0
\label{eq:chi2_2}
\end{equation}
equivalent to
\begin{equation}
\frac{\sqrt{\overline{\lambda}\,P_j}}{P_j}\ =\ \frac{I_1(\sqrt{\overline{\lambda}\,P_j})}{I_o(\sqrt{\overline{\lambda}\,P_j})}
\label{eq:chi2_3}
\end{equation}
where $I_n$ is the modified Bessel function of the first kind and index $n$.
At $P_j\gg 1$ the solution is $\overline{\lambda}=P_j$. Numerical solutions to
equation~\ref{eq:chi2_3} are displayed in Figure~\ref{f:lambda} (solid line),
which shows how rapidly $\overline{\lambda}$ converges to $P_j$ as a function of $P_j$.

In the $j=N/2$ case it is $r=1$ and the interested random variable is $\bmath{P}_{N/2}/2$.
To simplify the notation, let us define $x=P_{N/2}/2$, so it is
\begin{equation}
\chi^2_1(\lambda, x)\ =\ \frac{1}{\sqrt{2\,\pi\,x}}\,
e^{-(x + \lambda)/2}\ \cosh{(\sqrt{x\,\lambda})}\quad .
\label{eq:chi1}
\end{equation}
When $0\le x\le 1$ equation~(\ref{eq:chi1}) monotonically decreases for $\lambda>0$,
so it is $\overline{\lambda}=0$. When $x>1$, the maximum is found analogously to
equation~(\ref{eq:chi2_2}), thus
\begin{equation}
\frac{\partial \chi^2_1(\lambda, x)}{\partial \lambda}\Big|_{\lambda=\overline{\lambda}}\ =\ 0
\label{eq:chi1_2}
\end{equation}
equivalent to
\begin{equation}
e^{2\,\sqrt{\overline{\lambda}\,x}}\ =\ \frac{1 + \sqrt{\overline{\lambda}/x}}{1 - \sqrt{\overline{\lambda}/x}}
\quad .
\label{eq:chi1_3}
\end{equation}
Analogously to the $j<N/2$ case, the solution is $\overline{\lambda}\la x$ and rapidly
converges to $\overline{\lambda}=x$, as shown in Figure~\ref{f:lambda} (dashed line).
%
% +++++++++++++ lambda ++++++++++++++ 
%++++++++++++++++++++++++++++++++++++++++++
%

Summing up, in both cases it is $\overline{\lambda}=0$ for $P_j\le 2$, and for $P_j>2$ it
asymptotically tends to $P_j$ ($P_{N/2}/2$) for $j<N/2$ ($j=N/2$).

In the process of estimating the variance of $P_j$, we conservatively assume
$\overline{\lambda}=P_j$ ($j<N/2$), and $\overline{\lambda}=P_{N/2}/2$ ($j=N/2$), so
\begin{eqnarray}
\displaystyle \overline{\lambda} = \left\{\begin{array}{lr}
\displaystyle P_j & \qquad (j<N/2)\\
\displaystyle P_{N/2}/2 & \qquad (j=N/2)\\
\end{array}
\right. \quad .
\label{eq:lambdatot}
\end{eqnarray}
By replacing equation~(\ref{eq:lambdatot}) into equations~(\ref{eq:var_p}, \ref{eq:var_pNy})
equation~(\ref{eq:var_p3}) is obtained.

%

%%%%%%%%%%%%%%%%%%%%%%%%%%%%%%%%%
% REFERENCES
%%%%%%%%%%%%%%%%%%%%%%%%%%%%%%%%%

\end{document}